\documentstyle[12pt]{article}
\newcommand{\ua}{\uparrow}
\newcommand{\da}{\downarrow}
\newcommand{\Pspins}{\ua \da - \da \ua }

\begin{document}
\title{\vspace*{-1.8cm}
Weak Radiative Decays of
Hyperons: \\ Quark Model and Nonlocality
\vspace{-0.3cm}}
\author{
{P. \.{Z}enczykowski}$^*$\\
{\em Institute of Nuclear Physics,
Krak\'ow, Poland}
\vspace{-0.3cm}}
\maketitle
\begin{abstract} 
\begin{footnotesize}
It is proved that symmetry structure of the
parity-violating amplitudes of 
weak radiative hyperon decays in
the vector-meson dominance (VMD) approach, and
the violation of Hara's theorem in particular, are also obtained
when direct coupling $e_q~ \bar{q}\gamma _{\mu }q A^{\mu }$
of photon to quarks is used in place of VMD (with calculations performed
in the limit of static quarks). 
Thus, violation of Hara's theorem 
in VMD-based models does not result from the lack of gauge invariance.
It is further shown that, in the static limit of the quark model,
the current-algebra
commutator term in the parity-violating amplitudes of nonleptonic hyperon decays
and the parity-violating $\Sigma ^+ \to p \gamma $ decay amplitude
are proportional to each other. As a result, 
 Hara's theorem may be satisfied in this limit
 if and only if the contribution from the
 current-algebra commutator in nonleptonic hyperon decays is zero.
Violation of Hara's theorem is traced back to the
nonlocality of quark model states in the static limit.
It is argued that the ensuing intrinsic baryon 
nonlocality does not have to be unphysical.
It is stressed that the
measurement of the $\Xi ^0 \to \Lambda \gamma $ asymmetry
will provide very important information
 concerning the presence or absence of nonlocal features
in parity-violating photon coupling to baryons at vanishing photon momentum.
If the $\Xi ^0 \to \Lambda \gamma $ asymmetry is found negative,
Hara's theorem is satisfied but the gauge-invariant
quark model machinery 
predicting its violation must miss some contribution, or be modified. 
If experiment confirms positive  $\Xi ^0 \to \Lambda \gamma $ asymmetry, 
then, most likely,
Hara's theorem is violated.  Although positive 
$\Xi ^0 \to \Lambda \gamma $ asymmetry admits of the possibility that
Hara's theorem is satisfied, this alternative
is in disagreement with hints suggested by the similarity
of photon and vector-meson couplings and the
observed size of parity-violating nuclear forces.
\end{footnotesize}

\end{abstract}
$^*$ E-mail:zenczyko@iblis.ifj.edu.pl

\newpage

\section{Introduction}

Thanks to the experimental programmes pursued at Fermilab and CERN, 
important data on weak radiative hyperon decays (WRHDs) should be available
within the next year or two.  These data 
will provide crucial input, 
which will direct further attempts to 
understand WRHDs.

Theoretical approaches to WRHDs may be divided 
into two classes
according to whether they 
(1) satisfy or (2) violate the theorem due to Hara \cite{Hara},
which states that the parity-violating (p.v.) amplitude of the 
$\Sigma ^+ \to p \gamma $ decay should vanish in the SU(3) limit.
Since the assumptions of Hara's theorem are 
fundamental (ie. CP conservation, gauge-invariant local field theory at
hadron level, and the absence of exactly massless hadrons), 
one may be tempted to discard class (2), and
claim that any obtained
violation of the theorem must result from
an unjustified assumption or an
erroneous calculation.
Yet, despite the fundamental nature of 
the assumptions of Hara's theorem, 
the data seem to favour its
violation \cite{LZ}. Clearly,
it may happen that some of these data turn out erroneous, 
and that the combined set of WRHD data finally agrees
with the theorem.
However, irrespectively of
whether one rejects or admits of the
possibility that Hara's theorem may be violated in Nature,
it is important to identify the origins of the violation in any given 
calculation.
Only mathematically precise analyses may provide us with a deeper understanding
of the problem, and its resolution
 when crucial experimental data become available.

In the quark  model of Kamal and Riazuddin (KR) \cite{KR}, 
Hara's theorem is violated in the SU(3) limit. 
It has been claimed that this result is due to
 KR calculation being not gauge-invariant (see \cite{Hol2000} and
 references therein).
As pointed out in ref. \cite{ZenComm2} (see also \cite{Azimov}), such
 claims are based on logically incorrect inferences. 
Technically speaking, violation of Hara's theorem in KR originates from a
  contribution in which the intermediate quark enters its
mass shell, thus satisfying the condition for a regular free particle
\cite{Zen99}.  Since
quarks are not such particles, the KR calculation 
must be considered unphysical, {\em if taken literally}.  
On the other side, however, it
indicates
 what would be needed for the
violation of Hara's theorem to occur in Nature.
Since entering mass shell is
a feature of asymptotic states, the KR calculation
hints that the violation of Hara's theorem requires 
some kind of nonlocality. 
Connection with nonlocality can also be anticipated
from the example of ref. \cite{Zencounterexample}, in which it is proved 
at hadron level that the
violation of Hara's theorem may occur only if the (conserved) 
electromagnetic axial baryon
current exhibits some degree of nonlocality (see also \cite{ZenComm2}).

It may be argued that one should replace the KR approach 
with a model in which quarks are confined to a small region of space. 
Hadrons should be then well described by an effective local field
theory, and Hara's theorem {\em must} be satisfied as its 
violation then requires the presence of unobserved
 exactly massless hadron \cite{LZ}. 
The problem is, however, that quark unobservability may be 
taken care of only using models
based on what is {\em expected} of confinement, 
not on its calculable properties:
one cannot trace if and how quark unobservability modifies KR results.
This means that  the correctness of these
 expectations  cannot be proved or disproved.

Thus, we are stuck in
 a stalemate:
in order to resolve the puzzle, we have
to take quark unobservability into account.  Yet, we do not know how to do
that properly.
A possible way out 
consists in finding a description in which use of 
free or confined intermediate quarks is
 altogether  avoided.
Such reasoning 
led to the idea of the $SU(6)_W \times VMD$ approach of refs. 
\cite{LZ,Zen89,Zen91}, in which photons couple to hadrons
always through inter\-mediate vector mesons
(vector-meson dominance (VMD) 
is known to work extremely well).
Then, all unknowns related to quark-level problems
are hidden in the
p.v. meson-baryon couplings.
The merit of this approach is that the latter couplings 
are experimentally
accessible in nuclear p.v. processes. 
On the other hand, the use of VMD may be considered
questionable: ref. \cite{Hol2000} attributes
violation of Hara's theorem in
\cite{LZ,Zen89} to the lack of gauge
invariance. 
The question of gauge invariance arises
when VMD is understood in a dynamical sense with vector mesons
mediating the coupling. 
Still,
even if one rejects the KLZ scheme \cite{KLZ,Sakurai} 
ensuring gauge invariance of VMD, the explanation of 
 the violation of Hara's theorem in
 refs. \cite{LZ,Zen89} by gauge noninvariance 
 of the underlying calculations is incorrect.
 Proving this is one of the goals of the present paper.

In this paper (Sections 2,3) it 
is shown that the basic results of refs. \cite{LZ,Zen89} 
are {\em independent} of the above dynamical understanding of VMD
and hold also for manifestly gauge-invariant direct photon-quark coupling. 
Consequently, the violation of Hara's theorem obtained in
\cite{LZ,Zen89}
has nothing to do with gauge noninvariance. 

In Section 4 the whole scheme and, in particular, the pattern of WRHD
asymmetries are cross-checked against the
PCAC approach to NLHDs. It is 
shown that if the current algebra (CA) commutator in NLHDs is nonzero,
the approach leads to the violation of Hara's theorem, while 
predicting large positive asymmetry
of the $\Xi ^0 \to \Lambda \gamma $ decay.
It is stressed that negative $\Xi ^0 \to \Lambda \gamma $
asymmetry (automatically consistent with Hara's theorem)
would present a serious problem
for the quark model.
The decisive role of the sign of the $\Xi ^0 \to \Lambda \gamma $
asymmetry as far as the issue of the violation of Hara's
theorem is concerned can also be seen from Table \ref{tab1} where
present experimental branching ratios and asymmetries of WRHDs are
gathered together with the predictions of two typical approaches
to WRHDs.
\begin{table}
\label{tab1}
\caption{Asymmetries and branching ratios of four most important WRHDs: 
HS -Hara's theorem satisfied, 
ref.\cite{Orsay}; 
HV -Hara's theorem violated, 
ref.\cite{LZ}}
\begin{tabular}{lrrrrrr}
\hline
Decay & \multicolumn{3}{c}{Asymmetries} & \multicolumn{3}{c}{Branching
ratios$\cdot 10^3$}\\
&  experiment &  HS &  HV & 
 experiment & HS & HV \\
\hline
$\Sigma ^+ \rightarrow p \gamma$&$-0.76\pm0.08$&$-0.80^{+0.32}_{-0.19}$&
$-0.95$&$1.23\pm0.06$&$0.92^{+0.26}_{-0.14}$&$1.3-1.4$\\
$\Lambda \rightarrow n \gamma $&&$-0.49$&$+0.8$&
$1.75\pm0.15$&$0.62$&$1.4-1.7$\\
$\Xi ^0 \rightarrow \Lambda \gamma $&$+0.43\pm0.44$&$-0.78$&$+0.8$&
$1.06\pm0.16$&$3.0$&$0.9-1.0$\\
$\Xi ^0 \rightarrow \Sigma ^0 \gamma $&$-0.63\pm0.09$&$-0.96$&$-0.45$&
$3.34\pm 0.10$&$7.2$&$4.0-4.1$\\
\hline
\end{tabular}
\end{table}

Providing an interpretation for the origin 
of the violation of Hara's theorem
in the quark model constitutes our another aim.
In Section 5
we reflect on 
the concept of quark position.
This reveals 
 that for static quarks (relevant
 for the calculations in the SU(3) limit), the
 composite quark-model states 
  possess nonlocal quantum properties.
Thus, the violation of Hara's theorem 
is traced to intrinsic nonlocality of baryons.
The question whether this property of the quark model
constitutes a drawback or a virtue 
 is discussed and
 it is argued that in the limit of zero photon
momentum such nonlocality does not have to be unphysical.
It is stressed that the sign of the $\Xi ^0 \to \Lambda \gamma $
asymmetry will provide crucial information on whether  the
difficulties with WRHDs are due to a serious problem in
the quark model or to intrinsic baryon nonlocality.

In Section 6 we consider the issue whether a positive sign
of the $\Xi ^0 \to \Lambda \gamma $ asymmetry 
requires violation of Hara's theorem. It is pointed out
that the
connection between WRHDs and
the (observed) p.v. effects in $NN$
interactions constitutes an additional hint against Hara's theorem.

Our conclusions are given in Section 7.

\section{Quark model in the static limit}
In this section we discuss the essential 
steps of the derivation of our results
(as well as those of \cite{Zen89} 
and of paper \cite{DDH} by Desplanques, Donoghue and Holstein
(DDH)). 
It is known that explanation of the 
observed pattern
of the WRHD branching ratios  
requires that the dominant contribution should come from
$us \to du\gamma $ processes shown in Fig.1. 
Other possible diagrams
have been estimated in various papers as negligible \cite{LZ}.
We want to evaluate the joint effects of
the process of $W$-exchange and the direct coupling of
photon to quarks in an approach in which explicit
treatment of intermediate quarks as free
particles (present in KR) is avoided or at least confirmed 
in an independent way.
This was also the original motivation
for the VMD approach of ref. \cite{Zen89}. 
All complications from strong interactions reside either in the
wave functions of external baryons, or are 
taken into account by the completeness
of the set of states intermediate between photon emission and $W$-exchange.
Couplings of pseudoscalar and vector mesons are described by similar
diagrams with the photon replaced by a meson and a suitable choice of
current in the coupling. Although one may criticize the latter
 assumption on the
grounds that mesons are not pointlike, 
symmetry structure of these couplings 
suffices to derive the results of \cite{DDH}.
Since ultimately we are interested in the limit of zero mass difference between
initial and final baryons, 
our calculations shall be done in the extreme
nonrelativistic (static) limit, 
with momenta of both photon and individual quarks approaching zero. 
Taking this limit does not violate gauge invariance:
one might keep small terms of higher order in momenta 
and drop them at the end.
The validity of the assumption of static quarks
shall be discussed later.
Below we give a schematic 
presentation of quark model calculations. Practical
equivalence of $SU(6)_W$ and the quark model
technique was discussed in \cite{DDH}.

\subsection{Strong and electromagnetic couplings}
Parity-conserving (p.c.) interactions of fields with 
quark currents are given by:\\
 for pseudoscalar mesons
\begin{equation}
\label{pseudosc}
g_P~\bar{q}_l\gamma _5 q_m~P^{lm} 
\end{equation}
 for vector mesons
\begin{equation}
\label{vectorm}
g_V~\bar{q}_l\gamma _{\mu } q_m~V^{lm,\mu } 
\end{equation}
 for photons
\begin{equation}
\label{minimal}
e_l~\bar{q}_l\gamma _{\mu} q_l~A^{\mu } 
\end{equation}
where $l,m=u,d,s$ label quark-antiquark operator fields $q_l$. Colour
indices are suppressed.

In the static limit, neglecting inessential factors, one
may rewrite the relevant structures in 
Eqs.(\ref{pseudosc}-\ref{minimal}) as follows
(the direction of the 3rd axis is defined as the direction of 
meson/photon momentum ${\bf k}$
before taking the limit $k_{\mu } \to 0$):

for pseudoscalar mesons
\begin{equation}
\label{creationP}
[a^{+}(\bar{m}l(\Pspins))+a(\bar{l}m(\Pspins))]~
a^{+}(P(m\bar{l})) 
\end{equation}
(+ terms~not~involving~$a^{+}(P(m\bar{l}))$);

for vector mesons
\begin{equation}
\label{creationV}
-[a^{+}(\bar{m}l(\ua \ua))+ a(\bar{l}m(\da \da))]~
a^{+}(V(m\bar{l},-1))
\end{equation}
(+ terms~not~involving~$a^{+}(V(m\bar{l},-1))$);

for photons
\begin{equation}
\label{creationA}
-[a^{+}(\bar{l}l(\ua \ua))+ a(\bar{l}l(\da \da))]~
a^{+}(A(-1))
\end{equation}
(+ terms~not~involving~$a^{+}(A(-1))$).\\
Here $a^{+}(\bar{m}l(\ua\da -\da \ua))$ denotes difference of two
terms consisting of products of creation
operators of quark $l$ and antiquark $\bar{m}$ in spin states described
by arrows: 
\begin{equation}
a^{+}(\bar{m}l(\ua\da -\da \ua)) =
a^+(\bar{m}\ua)a^+(l\da )
-a^+(\bar{m}\da)a^+(l\ua )
\end{equation}
In $a(\bar{l}m(\Pspins))$ we have annihilation operators of
antiquark $\bar{l}$ and quark $m$. 

The ordering of indices corresponds to the ordering of quark
creation (annihilation) operators, which satisfy standard anticommutation
relations.
Furthermore, $a^{+}(P(m\bar{l}))$ is a creation operator of pseudoscalar field
describing meson composed of quark $m$ and antiquark $\bar{l}$, while 
$a^{+}(V(m\bar{l},-1))$ corresponds to vector meson with spin projection down.
Similarly,  $a^{+}(A(-1))$ describes creation of photon with
its spin directed along the negative axis.
For future discussion, we  note that for the coupling
\begin{equation}
\label{nonminimal1}
g'_V~\bar{q}_li\sigma _{\mu \nu}k^{\nu }q_m ~V^{lm,\mu }
\end{equation}
at small momentum $k^{\nu }$, $k^2 \approx 0$ one obtains 
the following structure ($k=k^0\approx k^3$)
\begin{equation}
\label{nonminimal2}
-k[a^{+}(\bar{m}l(\ua \ua))- a(\bar{l}m(\da \da))]~
a^{+}(V(m\bar{l},-1))
\end{equation}
The plus sign in Eq.(\ref{creationV}) 
is here replaced by the minus sign. 
Corresponding formulas
for photons are obtained from  
Eqs.(\ref{nonminimal1},\ref{nonminimal2}) by taking $l=m$ and performing
substitutions $V \to A$, $g'_V \to \kappa _l$, with $\kappa _l$ being the
anomalous magnetic moment of quark $q_l$. 
Formulas (\ref{creationP}-\ref{creationA},\ref{nonminimal2}) 
may be conveniently represented in terms of quark diagrams
shown below the dotted lines in Fig.1. 
The ordering of quark lines (from top to bottom) corresponds
to the ordering of quark creation (annihilation) operators.

\subsection{Weak interactions}
Starting from the standard $(V-A)\times (V-A)$ weak interaction,
after expressing  axial and vector weak currents relevant for 
the transition $us \to du$
in a way analogous to that given in the previous subsection,  
the contribution from the p.v.  part of
$W$-exchange is proportional in the static limit to
\begin{equation}
\label{Hpv}
H^{pv}_W =
H_u^{s,u\bar{d}}+H_d^{u,s\bar{u}}+H^s_{u,d\bar{u}}+H^u_{d,u\bar{s}}
\end{equation}
where

\begin{eqnarray}
H_u^{s,u\bar{d}}&=&
a^+(u\da)a(s\da )a(u\ua)a(\bar{d}\da)
-a^+(u\ua)a(s\ua )a(u\da)a(\bar{d}\ua)
\nonumber 
\\
     &+&          a^+(u\ua)a(s\da )a(u\ua)a(\bar{d}\ua)
		  -a^+(u\da)a(s\ua )a(u\da)a(\bar{d}\da)	
\end{eqnarray}
and
\begin{eqnarray}
H^s_{u,d\bar{u}}&=&a^+(u\ua)a(s\ua )a^+(d\da)a^+(\bar{u}\ua)
                  -a^+(u\da)a(s\da )a^+(d\ua)a^+(\bar{u}\da) \nonumber \\
&+&	         a^+(u\ua)a(s\da )a^+(d\da)a^+(\bar{u}\da)   
		 -a^+(u\da)a(s\ua )a^+(d\ua)a^+(\bar{u}\ua)
\end{eqnarray}
with $a^+(u)a(s)$ describing $s \to u$ transition of one of the quarks.
The above two expressions may be conveniently represented diagrammatically
as shown above the dotted lines in Fig.1. 
The ordering of three creation (or three annihilation) 
operators
corresponds to the ordering of relevant quark lines (from top to bottom).
Note that in this language
the description of p.v. processes is made possible by
the appearance of (negative parity) antiquarks \cite{DDH}.
Additional four terms with $d\leftrightarrow s$ are to be added to
Eq.(\ref{Hpv})
if $ud \to su$ processes are to be described as well. This makes the whole
interaction symmetric under $d\leftrightarrow s$.

Starting from the $(V-A) \times (V-A)$ interaction,
we have also found by explicit calculation
that the corresponding expression for $\bar{u}\bar{s}\to
\bar{d}\bar{u} $ is obtained from Eq. (\ref{Hpv}) by replacing quarks with
antiquarks and vice versa (without acting on the spin degrees of freedom), 
and by
changing the overall sign of the hamiltonian, as expected. 
Similarly, explicit calculation has shown that
the corresponding p.c. hamiltonian does not
change its sign after charge conjugation. These checks have confirmed
that the calculation is performed correctly and does not involve any
artificial CP-violating effects, 
forbidden by the assumptions of Hara's theorem.

\subsection{Final prescription}
The final calculation to be performed consists in the
evaluation of matrix elements of quark vector or pseudoscalar current 
by sandwiching it in between external baryonic states, described
by standard spin-flavour wave functions of ground-state baryons
and modified by
weak interaction of Eq.(\ref{Hpv}).

Calculations of the p.v. NLHD and WRHD amplitudes proceed
 in two steps:\\
(1) evaluation of the admixture of $q\bar{q}$ pairs in $qqq$ baryons
generated by Eq.(\ref{Hpv}), and\\
(2) calculation of the matrix elements of currents 
$\bar{q}\gamma _5 q$ and $\bar{q} \gamma _{\mu} q$ in between states
with these admixtures.\\
For p.v. WRHD amplitudes the 
matrix elements to be evaluated are of the form 
\begin{eqnarray}
&\langle q'_1 q'_2 q'_3 |~H_W^{pv}~ 
\bar{q}\gamma _{\mu} q~|q_1 q_2 q_3 \rangle &~~~~~~~~~~(b1)
 \nonumber\\
 \label{perturb}
&\langle q'_1 q'_2 q'_3 | 
~\bar{q}\gamma _{\mu} q ~H_W^{pv}~|q_1 q_2 q_3 \rangle
&~~~~~~~~~~(b2)
\end{eqnarray}
For NLHDs, replace the vector current $\bar{q}\gamma _{\mu} q$ 
with the pseudoscalar current $\bar{q}\gamma _{5} q$
(thus, NLHDs and WRHDs are related through the 
{\em spin properties of the quark model}).
Energy denominators have been suppressed:  
as they correspond to energy difference between $qqq$
and $qqq\bar{q}q$ states, they are identical for diagrams
(b1) and (b2).
The diagrammatic notation of Fig.1 helps in
the calculations which are a little tedious but straightforward.
They may be simplified by exploiting
total symmetry of ground-state baryon spin-flavour wave function.
Thanks to the symmetry of the wave functions,
it is sufficient to evaluate the contribution when photon (meson) 
emission proceeds from quark creation or annihilation operator
(from the $\bar{q}\gamma _{\mu} q$ current)
contracted with the third (by definition) quark in a baryon.  
The third quark is also one of quarks between which 
$W$-exchange occurs. Because of wave-function symmetry 
it is sufficient to
consider contributions in which the other quark undergoing weak interactions
is quark number two. The resulting diagrams to be evaluated are 
precisely the diagrams of Fig.1, in which quarks number 1,2,3 
are ordered
from top to bottom.
When the actions of weak hamiltonian on external 
states are worked out, one obtains:
\begin{eqnarray}
&\langle q_1 q_2 q' \bar{q'} q'_3 |~ 
\bar{q}\gamma _{\mu} q~|q_1 q_2 q_3 \rangle & ~~~~~~~~~~(b1) \nonumber \\
\label{5qto3q}
&\langle q_1 q_2 q'_3 | 
~\bar{q}\gamma _{\mu} q ~|q_1 q_2 q' \bar{q'} q_3 \rangle &~~~~~~~~~~(b2)
\end{eqnarray}
with $\bar{q} \gamma _{\mu} q $ acting on the bottom quark line, as shown
in Fig. 1.

The above equations are 
completely analogous to those appearing in the
standard quark-model calculation of
baryon magnetic moments: 
the latter are evaluated from the
nonrelativistic reduction of
\begin{equation}
\label{MM}
\langle q_1 q_2 q_3| ~\bar{q}\gamma _{\mu} q ~| q_1 q_2 q_3\rangle
\end{equation}  
where, because of the spin-flavour symmetry of external three-quark states,
contribution from the third quark only needs to be calculated.

\section{Pattern of parity-violating amplitudes and asymmetries}

 Using direct photon-quark coupling of 
Eqs. (\ref{minimal},\ref{creationA}) 
and the weak hamiltonian of Eq. (\ref{Hpv}), it is straightforward to
evaluate the  p.v.
WRHD amplitudes.
In this calculation, relative signs of various contributions are fixed by
the employed group-theoretic structure.
In particular, for the $\Sigma ^+ \to p \gamma $ p.v. amplitude one obtains
\begin{equation}
\label{Sigma}
A(\Sigma ^+ \to p \gamma) = -\frac{1}{3\sqrt{2}}b_{\gamma}
-\frac{1}{3\sqrt{2}}b_{\gamma}
\end{equation}
where the first term comes from diagram (b1) 
and the other one from diagram (b2).
The overall size of the amplitude is taken care of by $b_{\gamma }$, which
is proportional to electric charge $e$ and Fermi coupling constant $G_F$.
The subscript $_\gamma$ stresses that the calculation has been performed 
using direct 
photon-quark coupling, ie. without any intermediate vector meson.
Normalisation of numerical
factors in front of $b_{\gamma }$ has been chosen so that when the subscript
$_\gamma$ is omitted, one recovers formulas of Table 7.2 in ref.\cite{LZ} 
(Table 3 in ref. \cite{Zen99}). These formulas are repeated here in Table
2.
 Relative sizes and signs of all contributions
in these tables are a consequence of the group-theoretic assumption
of the quark model
 and {\em not} of the assumption of VMD in its dynamical sense.
This should be obvious from the comparison of 
interactions of vector meson and photons
with quarks in 
Eqs.(\ref{creationV},\ref{creationA}), where  vector currents
$\sum_l e_l \bar{q}_l\gamma _{\mu} q_l$ and 
$\sum _{l,m}  
g_V \bar{q}_l\gamma _{\mu} (\lambda_3+\lambda _8/\sqrt{3})_{lm}q_m$ 
are proportional. 
Clearly, the evaluation of matrix elements of a current in between hadronic
states is
completely oblivious to
the nature of the field coupled to it.
With quark currents for photon and vector-meson couplings  identical,
the calculated symmetry structure of  
p.v. couplings of vector mesons
and photons must be the same.
The only difference is the size of the coupling.
Thus, VMD may be also understood as a two-step 
{\em merely technical}  prescription of substitution:\\
i) - evaluate the couplings of vector mesons to 
hadrons in the quark model,\\
ii) - in order to see what direct photon-quark couplig would give,
perform the {\em substitution}
(here for the $\rho $ meson): 
$\rho _{\mu } \to \frac{e}{g_{\rho}}A_{\mu }$.\\
In this sense, the results of VMD
calculations of refs. \cite{LZ,Zen89,Zen91} 
cannot be any less gauge-invariant
than those obtained from minimal direct photon-quark coupling.

From Eq.(\ref{Sigma}) it follows that the p.v. amplitude 
for the $\Sigma ^+ \to p \gamma $ decay is equal to 
$-\frac{2}{3\sqrt{2}}b_{\gamma }$.
 Hara's theorem might be satisfied in the
SU(3) limit provided $b_{\gamma }$ vanishes.  
The latter is not true in our quark model calculations in the static limit.

\begin{table}
\label{b1b2}
\caption{Amplitudes $b_1$ and $b_2$}
\begin{center}
\begin{tabular}{ccc}
\hline
Decay & Diagram (b1) & Diagram (b2)\\
\hline
$\Sigma ^+ \to p \gamma $         & 
$-\frac{1}{3\sqrt{2}}b_{\gamma }$ & 
$-\frac{1}{3\sqrt{2}}b_{\gamma }$ \\
$\Lambda \to n \gamma $           & 
$+\frac{1}{6\sqrt{3}}b_{\gamma }$ &
$+\frac{1}{2\sqrt{3}}b_{\gamma }$ \\
$\Xi ^0 \to \Lambda \gamma $      & 
$0$                               & 
$-\frac{1}{3\sqrt{3}}b_{\gamma }$ \\
$\Xi ^0 \to \Sigma ^0 \gamma $    & 
$\frac{1}{3}b_{\gamma }$          & 
$0$                               \\
\hline
\end{tabular}
\end{center}
\end{table}

P.v. amplitudes of the remaining WRHDs 
(Table 2
) are also
proportional to $b_{\gamma }$.
This means that in order for Hara's theorem to be satisfied,
p. v. amplitudes of {\em all} WRHDs must vanish in the SU(3)
limit\cite{Zen99a}. 
Modifications of the predictions of the 
static limit  so as to take into account SU(3) breaking
 shall be discussed in Section 6.
In Sections 3 - 5 
we accept that experimental asymmetries of WRHDs at $k_{\mu} \ne 0$
can be well approximated by the static quark model prescription. 
A similar assumption was used in ref.\cite{DDH} when considering
p.v. couplings of  mesons to baryons.
Further discussion of this assumption shall be given later.

With the coupling of photon to quark vector current proportional to
the coupling of {\em U}-spin-singlet vector meson to that current, 
it is obvious 
that all the relative signs of p.v.
amplitudes calculated in the present
scheme with direct 
photon-quark coupling must  be proportional to the
amplitudes of the $SU(6)_W$ + VMD approach of \cite{LZ,Zen89}, ie. they
are given by the sums of entries in columns (b1) and (b2) 
in Table 2.
(For clarity and in order to exhibit some of
the $s \leftrightarrow d$ 
symmetry properties, 
the contributions from vector currents with
definite quark content are given
in Table 3 
for the $\Sigma ^+ \to p$ transitions with 
p.v. weak interaction in the initial or final baryon.)

\begin{table}
\caption{Weights of amplitudes $b_1$ and $b_2$ for $\Sigma ^+ 
\stackrel{\bar{q}\gamma _{\mu} q}{\longrightarrow} p $ in the presence of
strangeness-changing p.v. weak interaction ($W$-exchange)
in initial or final baryon
for vector currents with
well-defined $q\bar{q}$ content}
\begin{center}
\begin{tabular}{ccc}
\hline
Current  & Diagram (b1) & Diagram (b2) \\ $\bar{q}\gamma _{\mu} q$
&$\langle \Sigma ^+ | ~\bar{q}\gamma _{\mu} q ~H_W^{pv} ~|p\rangle $
&$\langle \Sigma ^+ | ~H_W^{pv} ~\bar{q}\gamma _{\mu} q ~|p\rangle $\\
\hline
$\bar{u}\gamma _{\mu} u$   &
$-\frac{1}{3\sqrt{2}}$          & $-\frac{1}{3\sqrt{2}}$         \\
$\bar{d}\gamma _{\mu} d$   &
$+\frac{1}{3\sqrt{2}}$          & $0$                            \\
$\bar{s}\gamma _{\mu} s$   &
$0$                             & $+\frac{1}{3\sqrt{2}}$         \\
$(2\bar{u}\gamma _{\mu} u-\bar{d}\gamma _{\mu}d-
\bar{s}\gamma _{\mu }s)/\sqrt{6}$       &
$-\frac{1}{2\sqrt{3}}$           & $-\frac{1}{2\sqrt{3}}$          \\
\hline
\end{tabular}
\end{center}
\end{table}

Description of WRHDs requires knowledge of 
p.c.
amplitudes as well. These amplitudes have been evaluated in many approaches,
such as different versions of the pole or quark models. In the pole model
the dominant contribution comes from intermediate ground
states. 
One may also identify a correspondence between the pole and quark model
prescriptions: up to some details, the symmetry structure of the
p.c. WRHD amplitudes is similar in both approaches. 
Therefore, p.c. WRHD amplitudes may be safely described
in terms of the pole model. Reliability of the model is confirmed
by its successful description of 
p.c. NLHD amplitudes (cf. \cite{LZ}).

From the pole model we know the approximate sizes and signs 
of the p.c. WRHD amplitudes.  In our conventions, these signs are 
$-,-,+,+$ for
 $\Sigma ^+ \to p \gamma $, $\Lambda \to n \gamma $, 
$\Xi ^0 \to \Lambda \gamma $, and $\Xi ^0 \to \Sigma ^0 \gamma $ respectively, 
see eg. \cite{LZ}. 
Upon taking the sum of contributions from diagrams (b1) and
(b2) (Table 2), the asymmetries of the
$\Sigma ^+ \to p \gamma $ and $\Xi ^0 \to \Sigma ^0 \gamma $ 
decays turn out to be
 of the
same sign, while for $\Xi ^0 \to \Lambda \gamma $ the asymmetry
is of the opposite sign \cite{LZ,Zen89}. 
Relative sizes of p.v. and p.c. amplitudes are such that if one of the
asymmetries is large, all of them are large.
In particular, in ref.\cite{LZ} 
the asymmetry in the $\Xi ^0 \to \Sigma ^0 \gamma $ decay
was predicted to be around $-0.45$ (or slightly more negative, see
\cite{ZenHyp99b}), 
away from its first
measurement of $+0.20 \pm 0.32$ \cite{Teige}.
A recent experiment \cite{KTeV2000} gave
$\alpha (\Xi ^0 \to \Sigma ^0 \gamma) =-0.63 \pm 0.09$,
confirming that the
$\Xi ^0 \to \Sigma ^0 \gamma $ asymmetry is significantly
negative 
(and pointing out an error in ref. \cite{Teige}).
With substantially negative 
asymmetry of $\Xi ^0 \to \Sigma ^0 \gamma $, 
using the experimental
branching ratio and the pole model prediction for the p.c.
amplitude of this decay,
one can determine $b_{\gamma }$ 
and predict the asymmetry of the $\Xi ^0 \to \Lambda \gamma $ to 
be large and positive. 
Although in ref. \cite{LZ} this prediction was obtained in the framework of
a VMD approach, {\em no dynamical VMD is needed to obtain this result}.  Here
this conclusion follows from direct photon-quark coupling
in the static limit of the quark model.

In fact, the sizes and relative signs of p.v. and p.c.
 WRHD amplitudes
can be fixed 
without any recourse
to experimental data on WRHDs or vector meson 
couplings to baryons.
One needs to know the data on NLHDs only.
Namely, in both p.v. and p.c. NLHD amplitudes there occurs the same
interaction of quark pseudoscalar current 
with an emitted pseudoscalar meson,
and, similarly, in both p.v. and p.c. WRHD amplitudes 
there occurs the same
 interaction of quark vector current 
 with photon.
 With spin-flavour symmetry providing the connection
 between the two currents, 
 the NLHD and WRHD $W$-exchange contributions to
  amplitudes become related (separately in p.c.
 and p.v. sectors)
 and the prediction for WRHD asymmetries and branching ratios becomes absolute. 
 Roughly speaking, one has to remove the factor of $g_P$ 
 from the NLHD amplitudes, perform the
 appropriate
 symmetry transformation (ie. 
 replace the pseudoscalar current with the vector one),
  and multiply the results by the electric charge.
 Up to some details, this
 prediction is numerically the one given in 
 refs. \cite{LZ,Zen89,Zen91}.
 Further discussion of the connection with NLHDs is given in the next section.

At this point one may ask what part of the results of \cite{LZ,Zen89,Zen91}
really depends on the assumption of VMD understood in its dynamical sense. 
The answer is: not much.
Since formulas for the p.v. amplitudes in the present scheme 
and in refs. \cite{LZ,Zen89,Zen91} are identical in the SU(3) limit, 
the only place where dynamical VMD effects 
may be somewhat important is the 
description of p.c. amplitudes. In fact, in
refs. \cite{LZ,Zen91} these amplitudes
are assumed in a form  obtained  from
 the experimental p.c. NLHD amplitudes
by symmetry transformation from pseudoscalar to vector mesons.
This amounts to some fine tuning of the p.c. WRHD amplitudes as compared
to calculations avoiding that route, thus helping a little with the fits. 
The signs and approximate size of asymmetries in \cite{LZ,Zen91} 
do not depend on this fine tuning, however.
 
\section{Relative sign of $b_1$ and $b_2$ amplitudes and PCAC}

Although the direct photon-quark coupling approach 
gives absolute predictions
for the pattern of asymmetries in the static limit, 
 an independent check on its predictions would be welcome, 
especially in view of the fact that
the pattern depends on one single relative sign
between the contributions from diagrams (b1) and (b2). 
One may wonder whether this
sign has been fixed correctly or whether
the intermediate states have been treated in a way consistent
with experiment elsewhere. 
To answer these questions, we turn to the 
PCAC reduction of p.v. NLHD amplitudes in the limit
when pion momentum $k_{\mu }$ goes to zero.

Direct calculation of the contribution from $W$-exchange processes to the
p.v. NLHD amplitudes
along the lines of Section 2 gives the results shown in
Table \ref{tabNLHD}. The amplitude  $b$ in Table \ref{tabNLHD} 
is proportional
to $b_{\gamma }$ used in WRHDs in the previous section, 
with the proportionality factor 
including  $g_P/e$ (compare Eqs.(\ref{pseudosc},\ref{minimal})).
\begin{table}
\label{tabNLHD}
\caption{$W$-exchange-induced ($H^{pv}_W$) 
contributions to p.v. amplitudes
and  the corresponding expressions obtained through PCAC reduction}
\begin{center}
\begin{tabular}{c|cc|c|cc|c}

\hline
&\multicolumn{3}{c|}{$\langle p P|H^{pv}_W | \Sigma ^+\rangle $}
&\multicolumn{3}{c}{$\langle \Sigma ^+ P|H^{pv}_W | p \rangle $}\\
meson P ($q\bar{q}$)& $(b1)$ & $(b2)$ & PCAC &$(b1)$ & $(b2)$ &PCAC \\
\hline
$u\bar{u}$ &$-\frac{1}{2}b$ &$\frac{1}{2}b$&&
$-\frac{1}{2}b$&$\frac{1}{2}b$&\\
$d\bar{d}$ &$-\frac{1}{2}b$ &$0$&&
$0$&$\frac{1}{2}b$&\\
$s\bar{s}$ &$0$ &$\frac{1}{2}b$&&
$-\frac{1}{2}b$&$0$&\\
$\frac{1}{\sqrt{2}}(u\bar{u}-d\bar{d})$ &
$0$ &$\frac{1}{2\sqrt{2}}b$& 
$\frac{i}{2F_{\pi }}\langle p|H^{pc}| \Sigma ^+ \rangle $ 
&$-\frac{1}{2\sqrt{2}}b$&$0$& 
$-\frac{i}{2F_{\pi }}\langle \Sigma ^+ |H^{pc}| p \rangle $
\\
$\frac{1}{\sqrt{6}}(2u\bar{u}-d\bar{d}-s\bar{s})$ &
$-\frac{1}{2\sqrt{6}}b$ &$\frac{1}{2\sqrt{6}}b$& $0$&
$-\frac{1}{2\sqrt{6}}b$ &$\frac{1}{2\sqrt{6}}b$&$0$\\
\hline
\end{tabular}
\end{center}
\end{table}

From the comparison of the prediction of the static quark model 
with the results of the
PCAC calculation (also given in Table \ref{tabNLHD}),
we see that the relative sign of contributions from diagrams
(b1) and (b2) has to be positive.
Indeed, on account of $d\leftrightarrow s $ symmetry of weak
hamiltonian we have
\begin{equation}
\langle p|H^{pc}| \Sigma ^+ \rangle =
\langle \Sigma ^+ |H^{pc}| p \rangle 
\end{equation}
and a negative sign between contributions (b1) and (b2) 
would lead to contradiction with the PCAC description
of p.v. NLHD amplitudes.
This is also seen when $P$ is a {\em U}-spin singlet
$\frac{1}{2}[\pi ^0+\sqrt{3} \eta _8]$.
Please note that the formulas obtained by PCAC reduction 
do {\em not} depend on any intermediate states.
Thus, for NLHDs the relative positive sign between the two contributions
 has been confirmed for $k_{\mu } \to 0$ 
 in a way that {\em completely avoids using any
intermediate states}. 
Since within the calculational scheme employed,
the difference between NLHDs and WRHDs consists solely in the replacement of
the quark pseudoscalar current by the vector one,
the relative contributions from diagrams (b1) and (b2) 
in the p.v. WRHD amplitudes (Table 2)
must be added. In other words,
consistency with formulas obtained 
by PCAC reduction of NLHD
amplitudes in the $k_{\mu }\to 0$ limit
requires that in this limit the contributions from diagrams (b1) and (b2) 
have to be
added for the $\Sigma ^+ \to p \gamma $ p.v. amplitude
as well (as in Eq.(\ref{Sigma})):  
the $k^{\mu} \to 0$ limit is
relevant for both current-algebra (CA) 
soft-pion estimates and for Hara's theorem.
It follows that consistency with the PCAC-based description of NLHDs  
requires in particular that
the $\Xi ^0 \to \Lambda \gamma $ asymmetry be positive, in disagreement with
ref. \cite{Orsay} and  chiral approaches \cite{Zen00}.

At this point it is appropriate to discuss the connection between 
the framework considered in this paper so far and 
the standard 
CA prescription supplemented with SU(3)-breaking 
 resonance-induced corrections.
Namely, the p.v. NLHD amplitudes 
$A(B_i \to B_f \pi )$ are usually
described in terms of the contribution from the CA commutator
plus terms proportional to pion momentum $k_{\mu }$:
\begin{equation}
\label{CAplusSU3break}
A(B_i \to B_f \pi ^a) = -\frac{i}{F_{\pi }} 
\langle B_f|[F^a_5,H^{pv}]|B_i\rangle+ k_{\mu}M^{\mu }
\end{equation}
The second term on the right vanishes in the SU(3) limit $k_{\mu } \to 0$.
This term is thought to be dominated by 
the SU(3)-breaking corrections from the intermediate 
$({\bf 70},1^-)$ resonances \cite{Orsay0}.
Using the properties of the Cabibbo hamiltonian 
to replace 
 $[F^a_5,H^{pv}]$ with  $[F^a,H^{pc}]$, and 
evaluating the contribution from resonances,
one obtains:
\begin{equation}
\label{reduced}
A(B_i \to B_f \pi ^a) \propto 
\langle \tilde{B}_f |H^{pc}| \tilde{B}_i \rangle + 
const( B_f,B_i) \times (m_s-m_d) 
\end{equation}
where $\tilde{B}_{i(f)}$ denote baryon states, of which one
is obtained from $B_{i(f)}$ by
the action of isospin generator $I^a$ and the other is left unchanged.
The leading term (in SU(3) breaking) in Eq.(\ref{reduced}) is 
$\langle \tilde{B}_f |H^{pc}| \tilde{B}_i \rangle$.
It contains the $W$-exchange-induced term 
$\langle \tilde{B}_f |H^{pc}_W| \tilde{B}_i \rangle$.
It must correspond to the $b_1+b_2$ term obtained in 
the scheme of this paper  at $k^{\mu } \to 0$:
one might have assumed that the states $|B_i\rangle , |B_f\rangle $ 
in between which  the CA commutator is to be
evaluated are the states of the static quark model.

Alternatively, one might {\em saturate} the CA commutator
with this part of contribution from resonances which does {\em not} 
vanish in the SU(3) limit.
That is, one may replace the r.h.s of Eq.(\ref{CAplusSU3break}) with the
pole model prescription. This amounts to considering energy denominators
for the intermediate states in the amplitudes of diagrams (b1) and (b2).
When one restricts to the $({\bf 70},1^-)$ ($B^*$) resonances 
in intermediate states,
for $\Sigma \to N $ transitions these energy denominators are 
$m_{\Sigma }-m_{N^*}$ for diagram (b1) , and
$m_N-m_{\Sigma ^*}$ for diagram (b2).
Assuming 
\begin{eqnarray}
m_{\Sigma} & \approx & m_N+(m_s-m_d) \nonumber \\
m_{\Sigma ^*} & \approx & m_{N^*}+(m_s-m_d) \nonumber \\
m_{N^*}-m_N = &\omega & = m_{\Sigma ^*}-m_{\Sigma } 
\end{eqnarray} 
we see that $(m_s-m_d)$ enters
with opposite signs into the two denominators. Thus SU(3) corrections to
diagrams (b1) and (b2) are of opposite signs:
\begin{equation}
\label{b1minusb2}b_1+b_2 
\stackrel{SU(3)~breaking}{-\!\!\!-\!\!\!-\!\!\!\longrightarrow}
\frac{b_1\omega}{\omega -(m_s-m_d)}+\frac{b_2\omega}{\omega + (m_s-m_d)}
\approx
b_1+b_2 + (b_1-b_2)\frac{m_s-m_d}{\omega}
\end{equation}
As a result, the symmetry structure of the SU(3)-breaking correction 
in Eq. (\ref{reduced}) is different from
the SU(3)-preserving CA term ($b_1-b_2$ versus $b_1+b_2$). 
Please compare the form of the 
right-hand sides of Eqs.(\ref{reduced}) and (\ref{b1minusb2}).

In the approach to WRHDs discussed in \cite{Orsay} 
and in chiral perturbation theory (ChPT),
only the WRHD counterpart of the $b_1-b_2$ term above  is considered.  
In that approach there is no counterpart to
the CA commutator term of NLHDs: the $b_1+b_2$ term in WRHDs
is set to zero by hand because it is thought to violate gauge invariance.
On the other hand, the gauge-invariant scheme of this paper
indicates that such a term 
(generated by the replacement of current $\bar{q}\gamma _5 q$ for NLHDs with
current $\bar{q}\gamma _{\mu} q$ for WRHDs)
exists
and  has symmetry properties corresponding to the sums
of  $b_1$ and $b_2$ (Table 2).
Therein lies the difference between the genuine quark model approach
and the standard approach of ref. \cite{Orsay} or more recent ChPT
attempts.

With  nonzero $b_{\gamma }$ it follows that
Hara's theorem must be violated, unless a good reason for rejecting
the WRHD counterpart to the CA commutator is given.  
One cannot claim that this counterpart to the CA commutator violates gauge
invariance:
the input into the calculation is gauge-invariant direct
photon-quark coupling
and all the steps in the calculation are correct.  
The (nonzero) $b_1+b_2$ term  must be present in NLHDs because it corresponds
to  nonzero value of $\langle p | H^{pc}| \Sigma ^+ \rangle 
= \langle \Sigma ^+ | H^{pc} | p \rangle $ in the quark model in the
SU(3) limit.
This is in agreement with the present descriptions of NLHDs,
in which the contribution from the commutator is nonzero and large.
Subtracting the $b_1+b_2$ term 
in WRHDs while keeping it in NLHDs is completely arbitrary.
Thus, the $b_1+b_2$ term 
should be present in both NLHDs (the CA commutator) and
 WRHDs (the counterpart to the CA commutator).
 
 As the p.v. $\Xi ^0 \to \Lambda \gamma $ amplitude is due to amplitude
 $b_2$ only (Table 2),
 a negative sign of the $\Xi ^0 \to \Lambda \gamma $
 experimental asymmetry would mean 
 that amplitudes $b_2$ must enter with an additional
 negative sign, thus signalling the $b_1-b_2$ structure,
 and the cancellation of $b_1$ and $b_2$ terms for
$\Sigma ^+ \to p \gamma $.  However, what is more important,
 a large negative $\Xi ^0 \to \Lambda \gamma $ asymmetry would also
mean that either something 
is badly wrong with the quark-model machinery which
connects the calculations for  quark pseudoscalar and vector currents
(I reject this possibility because it concerns {\em spin} 
properties of the quark
model),
or a completely new contribution reveals itself in WRHDs {\em only},  not
only
cancelling the counterpart to the CA commutator, 
but effectively reversing its sign. 
There is no hint of
what such a contribution could be.

As mentioned in Section 2,
calculations of the p.v. amplitudes proceed
in two steps:
(1) evaluation of the admixture of $q\bar{q}$ pairs in $qqq$ baryons, and
(2) calculation of the matrix elements of currents 
$\bar{q}\gamma _5 q$ and $\bar{q} \gamma _{\mu} q$ in between states
with these admixtures.
As the latter do not depend on the current, there is
no reason why the scheme should fail for the vector current while being correct
for the pseudoscalar one.
Thus, if the CA commutator is dominant in NLHDs 
(the $k_{\mu } \to 0$ term dominates),
positive $\Xi ^0 \to \Lambda \gamma $ asymmetry is a {\em must}.
Consequently, negative $\Xi ^0 \to \Lambda \gamma $ experimental asymmetry 
would signal deep trouble for the quark model.
Note also that the violation of Hara's theorem 
must be related to step (1) above, ie.
to the properties of states in the quark model (the interaction
$\bar{q} \gamma _{\mu} q A^{\mu }$ of step (2) is gauge invariant).

\section{Origin of the violation of Hara's theorem in the quark model}

As discussed, 
the static gauge-invariant quark-model calculation 
unavoidably violates
Hara's theorem. 
This is in direct
conflict with what is easily proved in hadron-level approach. 
Before we jump to the conclusion
that there is something wrong with the static quark model, let us 
 discuss the origin of our
result from a different perspective.

We start with the standard hadron-level picture.
In this picture hadrons are assumed
to be well described by an effective local field theory.
We are interested in what happens at  point $k_{\mu } = 0$.
This point 
(as any other point in momentum space) 
corresponds to particles (baryons, photons)
 described by plane waves. The positions of
corresponding particles are not well defined: the particles may be found
anywhere. It might help to
think of particles as "potentially being everywhere".

The static-limit
calculations in the quark model involve juggling
spin, flavour and parity (quark-antiquark) indices only
(see Eqs.(\ref{creationA},\ref{Hpv},\ref{perturb},\ref{5qto3q}), Fig.1):
the quarks themselves are in states of vanishing
but definite momenta.
This prescription,
when {\em interpreted} in position space,
corresponds to the situation in which positions of
quarks "are" arbitrary, ie. now 
it is {\em individual} quarks that 
"are potentially everywhere".
In particular, this includes configurations
with baryon quarks 
 "potentially being" arbitrarily far away from each other.
Such nonlocal quark configurations  are forbidden in an effective
hadron-level local field theory, when quarks are assumed
to be close to one another.
Thus, the static quark model includes
configurations which are not and cannot be taken into account
in the language of effective hadron-level local field theory.
Hadron- and quark- level prescriptions are generically {\em different}.
Quark-model violation of Hara's theorem 
must come from these nonlocal configurations. 

The nonlocal configurations  of  baryon quarks
 may be considered unacceptable: their presence is
 in conflict with the expectations based on the idea of
confinement. 
We know, however, that in general
composite quantum states 
may exhibit nonlocal features. 
Since baryons are quantum states made out of quarks, 
the real question is whether we
can exclude such (admittedly weird)
configurations on the grounds of either experiment 
or of general 
ideas of quantum physics
(as opposed to those based on certain theoretical expectations)? 
Consider therefore the situation from the
point of view of what is measurable.  
 The closer we approach the
  $k_{\mu} = 0$ limit
 (assuming we can manipulate $m_s-m_d$),
 the worse spatial resolution we have.
 Consequently, we cannot  
 experimentally exclude that in the $k_{\mu}=0$
 limit, quarks may be thought of as "being" arbitrarily 
far away from each other, ie. that the photon-baryon coupling 
 is intrinsically nonlocal. 
Note that if experiment
 is set up to measure momenta
 it is not meaningful to talk
 about hadron position in any other way than with the help of a theory.
 The same applies to the positions of quarks.
In order to see small distances and
check what quark positions "really are", one has to
look "deep into hadrons". 
This requires high, not low momenta and amounts to asking a completely 
different experimental question. Consequently, 
 it is hard to see how one can reject the violation of Hara's theorem
 on the grounds of general principles.
 This interpretation 
of quark-model results 
 is clearly more general
than the model considered: 
the composite quark state may couple to zero-energy
photon in a 
genuinely nonlocal way.

 Note that the interpretation in terms of 
 quarks "being" arbitrarily far away naturally
 follows also when baryon magnetic moments are calculated 
 in the original quark model (cf Eq.(\ref{MM})).
 A possible resolution to this problem of free quarks, corresponding to 
  the {\em expected} effective local
 theory at hadron level, was proposed a long time ago in the form of
  the confinement idea.
 When confinement is imposed in a way consistent with the standard expectations, 
 no clear-cut conflict between naive quark model calculations 
 and hadron-level expectations emerges for baryon magnetic moments.
 For WRHDs, however, the
predictions of the static quark model, and those
of 
 the hadron-level effective local theory,
 are different.
 Thus, it appears that WRHDs probe in a subtle way
 the original question of apparent
 quark freedom and unobservability at low energies.
 For this reason, I think that the issue of the violation of
 Hara's theorem is extremely interesting and very important.

 There is another way to see that the violation of Hara's theorem requires
 some kind of nonlocality. 
 Namely, it has been shown in ref. \cite{Zencounterexample} that 
 when the violation of  Hara's theorem  
 with built-in current conservation is 
 {\em forced} into the "corset" of effective hadron-level
 local description, the electromagnetic axial baryonic
 current cannot be strongly suppressed at 
 infinity: the fall-off must be as slow as $1/r^3$.
 This should be contrasted with models in which the assumed exponential
 suppression of the current
 leads to Hara's theorem. 
 Since there is no 
 massless hadron,
the nonlocality discussed in ref. \cite{Zencounterexample} should be
thought of as simulating some other kind of nonlocality.

 One may argue that the static calculation is inadequate and
 that small components of Dirac spinors should be taken into account. 
 Indeed,
 small components necessarily emerge 
 when quarks are confined to a restricted
 volume of space. One may hope that contributions from
 these components could
 cancel the static limit term, 
 thus restoring Hara's theorem. This may be so 
 (see, however, ref.\cite{Lo}).
 The real problem is that one should include such terms
 in other calculations as well.
In particular, they should be taken into account in the calculations of
  p.v. meson-baryon couplings. 
  The relevant calculations (equivalent to those of Section 2)
  were performed by DDH \cite{DDH} 
  in the nonrelativistic (actually
  static) limit.  
 Thus, the celebrated DDH results stem from an approach
that violates
standard ideas about confinement.  

As the photon and vector-meson p.v.
couplings to baryons are calculated using the same quark-level current,
the Dirac structures of photon and vector-meson couplings
at hadron level should be identical. The only gauge-invariant coupling
acceptable in local hadron-level framework is that involving 
$\bar{B}_f \sigma _{\mu \nu}\gamma _5 k^{\nu} B_i$ (with $B_{i(f)}$ being baryon
bispinors). If Hara's theorem
is satisfied, also for vector mesons this structure only should appear.
However, the interaction
$\bar{B}_f \sigma _{\mu \nu}\gamma _5 k^{\nu} B_i V^{\mu }$ is inconsistent with
the data on p.v. effects in $NN$ interactions
(see Section 6).

 The above arguments suggest 
 that the language of effective local theories  at hadron level
 might sometimes constitute an insufficient approximation to 
 reality.
 In other words, the assumption that spin-1/2 baryon may {\em always} 
 be well
 approximated by a Dirac spinor field depending on a well-defined
 single point $x$ might be too 
strong\footnote{Field theories of nonlocal type have been
considered
as a possible theoretical vehicle for describing hadrons
since the times of Yukawa \cite{Yukawa}}. 
 Although one could 
 speculate about possible deeper origins and implications
 of the suspected limitations of the standard language, such 
 speculations will be warranted only if large
 positive value of the $\Xi ^0 \to \Lambda \gamma $ experimental
 asymmetry  is established.

 Of course, any limitation of the standard hadron-level language
  does not mean that the quark-level theory cannot be local 
or that individual quarks have to exhibit 
intrinsic nonlocality.
The calculation of
this paper constitutes an explicit counterexample.
Point-like interactions of the standard model underly the whole picture
herein considered.
It is the question of proper description of baryons as 
composite states that is 
being discussed here. Clearly, 
an effective local field theory
at hadron level should constitute a sufficient 
approximation for most of the present practical purposes.

\section{Dependence of $b_{\gamma }$ on SU(3) breaking parameter $m_s-m_d$}
So far,  the (mutually proportional) amplitudes 
$b_{\gamma }$ and $b$ have been nonzero constants. As the calculation was
performed in the SU(3) limit, these constants were independent of
the SU(3) breaking parameter $m_s-m_d$.  
This was also the 
case in ref. \cite{DDH}, 
where the $\Delta S =1$ p.v. amplitudes 
extracted from NLHDs (and in particular, their scale) 
were used to the description of $\Delta S =0$ 
nuclear p.v. processes and their scale.

Let us assume that experiment confirms the pattern 
corresponding to the sum of  $b_1$ and $b_2$,
thus indicating spin symmetry between NLHDs and WRHDs at $m_s \ne m_d$. 
This does not mean yet that the preceding sections describe
the physical situation in a qualitatively correct way.
The static quark model gives predictions at $k_{\mu }=0$, while in the real
world $k_{\mu } \ne 0$. 
Still, one suspects that
spin symmetry would
 survive should the masses of $s$ and $d$ quarks be different.
Thus, in a theoretical description in which quark masses are free parameters,
let $b \propto b_{\gamma }$ be a function of $\delta \equiv m_s-m_d$
(ie. $b\to b(\delta)$), which can be expanded into a series
 in the vicinity of $\delta =0$
  (thus, the
  form $b \propto b(0)+b'|m_s-m_d|$ is not allowed).
 Hara's theorem is satisfied if $b(0)=0 $. 
From symmetry properties under $s \leftrightarrow d $ interchange 
 (Table \ref{tabNLHD}) we see that only nonzero 
even powers of $m_s-m_d$ may appear in the expansion
of  $b(\delta )$ and $b_{\gamma}(\delta )$. 
With $b(0)=0$,
the lowest order term is proportional to $(m_s-m_d)^2$.
 If $b(\delta)= b'' \delta ^2$, the relevant hadron-level structure of 
photon-hadron p.v. interaction 
 may be written as 
\begin{equation}
\label{Harab1plusb2}
(m_s-m_d)(b_1''+b_2'')\bar{B_f}i\sigma _{\mu \nu}\gamma _5 k^{\nu } B_i A^{\mu }
\end{equation}
 where
 $b_1''(b_2'')$ denotes amplitude for diagram (b1)(diagram (b2))
 with the $(m_s-m_d)^2$ factor removed.
 In ref. \cite{Zen99a}, a pole model leading to the above solution
 was discussed. 
 Clearly, this way of making positive 
 $\Xi ^0 \to \Lambda \gamma $ asymmetry
 consistent with Hara's theorem
 is in conflict with the nonzero value of the CA commutator
 and with the quark model results for
 $\langle p |H^{pc}|\Sigma ^+ \rangle $, according to which 
the contribution of $W$-exchange between quarks 
does {\em not} vanish for $m_s=m_d$. 
 Internal consistency requires that for $m_s=m_d$ both
$\langle p |H^{pc}_W|\Sigma ^+ \rangle $ and $W$-exchange contributions to 
$A(\Sigma ^+ \to p \gamma )$ (Eq.(\ref{Sigma}))
behave in the same way in the static limit: they are either both zero or
both nonzero.

Note that if Hara's theorem is to be satisfied in the SU(3)
limit for positive 
 $\Xi ^0 \to \Lambda \gamma $ asymmetry , 
 {\em all} p.v. amplitudes must vanish in this limit 
 (Eq.(\ref{Harab1plusb2})).
 This is not the resolution proposed in 
\cite{Orsay} and chiral approaches 
\cite{Hol2000}, 
where only the $\Sigma ^+ \to p \gamma$
p.v. amplitude vanishes,
while the p.v. amplitudes of 
the remaining three WRHDs stay nonzero.
The symmetry structure of the latter approach yields the pattern $(-,-,-,-)$
for the asymmetries of $\Sigma ^+ \to p \gamma, \Lambda \to n \gamma, 
\Sigma ^0 \to \Lambda \gamma$, and
$\Sigma ^+ \to \Sigma ^+ \gamma $ respectively.
Such symmetry structure might be obtained
in the static limit with direct photon-quark coupling
if the photon were to couple to quarks through the nonminimal coupling of 
Eqs.(\ref{nonminimal1},\ref{nonminimal2}). This is 
 not acceptable.

 Extracting information
 on $b_{\gamma }(\delta) $ for small $\delta $ from the analysis 
 of NLHDs and WRHDs is impossible: with $m_s$ fixed,
there is no way to check whether the transition
amplitudes
are proportional to $(m_s-m_d)^2$ or not. 
Despite that,
 it is still possible to get some experimental indication as to
 which of the two theoretical solutions:  $b_{\gamma }(0) \approx
 b_{\gamma }(m_s-m_d)$ or $b_{\gamma }(0) = 0 $, is
 favored. 
 In order to see what these
 indications are, we recall that the coupling of photon to baryon 
 results
 from $\gamma $ coupling to quark vector current.
 As discussed in Section 2, the coupling of {\em U}-spin-singlet
 vector meson to baryons results
 from its coupling to the {\em same} current.
 Since the whole quark model machinery does not depend on the nature of
 the field this current is coupled to, we conclude that one should be able to
 get some hints
 about the possible dependence of $b_{\gamma }$ on $m_s-m_d$
 by looking at the $\Delta S =0$
 p.v. couplings of vector mesons to nucleons.
 If the factor of $(m_f-m_i)^2$ is present in $b$ and $b_{\gamma }$, 
 it follows
 that
 {\em all} $\Delta S =0$ p.v. weak couplings of mesons to baryons 
 should be very small. 
 Thus, we
 should observe nearly vanishing p.v. effects in the interactions
 of nucleons. 
 With respect to NLHD-based SU(3) estimates,
 they should be down by a factor of the order of at least
 $(m_n-m_p)^2/(m_N-m_{\Sigma})^2 \approx (m_u-m_d)^2/(m_s-m_d)^2 
 \approx 10^{-3}$ or less depending on the process 
 and the precise mass values used.
 This is not what is experimentally observed.
 The present data on p.v. nuclear forces
 indicate that the p.v. couplings of mesons
 to nucleons are roughly of the order expected from NLHDs 
 by an SU(3)-symmetric
 extrapolation, ie. as if there was no $(m_f-m_i)^2$
 suppression. This is in particular supported by 
 p.v. effects observed in $pp$ scattering, 
 which constitute the best experimentally
 established effect \cite{DDH,Hol2000b, Desplanques}.
 These data require 
 the presence of a $\overline{N}\gamma _{\mu}\gamma _5 N
 V^{\mu}$ term, and in particular, nonvanishing 
 contributions from $b$ diagrams. 
 If these $b$ (and the so-called $c$, cf. \cite{DDH}) diagrams 
 were negligible
 and only the factorisation $a$ terms (see \cite{DDH}) were allowed,
 one could not describe the p.v. effect observed in $pp$
 scattering \cite{Desplanques,Des95}.
 A $\overline{N}i\sigma _{\mu \nu}k^{\nu} \gamma _5 N V^{\mu}$ term
 cannot describe the data \cite{Desplanques} either.
 This suggests that the coupling of photons
 need not vanish, unless some additional cancellation takes place.
 This reasoning is correct, provided p.v.
 couplings of vector mesons to baryons can be correctly
 described by their evaluation in the static limit of the quark model.
 This is how they were estimated in \cite{DDH}, where they
 were also linked
 through symmetry to the soft-pion contribution to p.v.  
 NLHD amplitudes.

 It may be that the effects observed in nuclear parity violation  
 mean that the $\Delta S=0$ 
 p.v. couplings of vector mesons to baryons 
 cannot be reliably estimated in this way. For example it may be argued
 that in the p.v. nuclear force  the couplings of vector
 mesons at $k^2 = m^2_{\rho }$ should be used, while the scheme of this paper
 yields them at $k^2=0$.
 Clearly, experimental
 results on p.v. nuclear forces may be considered as a hint
 only.
 In the meantime, in order to learn more about WRHDs
 we have to wait for the results of the KTeV and NA48
 experiments on $\Xi ^0 \to \Lambda \gamma $ decays \cite{Ramberg99,Koch99}.
 
 \section{Conclusions}
 In this paper we have analysed the assumptions
 needed to generate the formulas of the VMD approach 
 to WRHDs \cite{Zen89,LZ}.
 We have shown that all their characteristic
 features 
 are obtained also in an explicitly gauge-invariant
  quark-level scheme in the limit of static quarks 
  with direct photon-quark
 coupling. 
 Consequently, 
 the claim of ref. \cite{Hol2000} that the results of 
 the VMD $\times $  $SU(6)_W$ scheme
 of refs. \cite{Zen89,LZ} (and  
 the violation of Hara's theorem) 
 follow from a lack of gauge invariance, is incorrect.
 
Our analysis shows that the standard pole-model approach of \cite{Orsay}
 and ChPT miss a quark-level
 contribution, which we call the WRHD counterpart to the CA
 commutator in NLHDs
 (the lowest-order term in ChPT). 
 This contribution is obtained by
  replacing the 
 interaction of pseudoscalar meson with the quark pseudoscalar current 
 by the interaction of photon with the quark vector
 current, ie. through the use of {\em spin properties of the quark model}.
 The obtained term is gauge invariant and
 proportional to matrix elements of the
 CA commutator in NLHDs. Consequently, the vanishing of the latter matrix
 elements  
 in between two states of strangeness differing by $\Delta S=1$
 should hold in the $m_s \to m_d$ limit if Hara's theorem is to 
 be satisfied.
 Because these matrix elements are proportional to
 the elements of the p.c. weak hamiltonian in between similar
 two baryonic states, the latter matrix elements should also vanish 
 in this limit. However, in quark model calculations
 they are non-zero.

 Detailed analysis of the direct photon-quark
 approach shows that in the quark model 
 the pattern of asymmetries predicted in refs.
 \cite{Zen91,LZ}
 is unavoidable.  
 In particular, large positive $\Xi ^0 \to \Lambda \gamma $
 asymmetry is necessarily predicted.
Any other experimental value for this asymmetry constitutes 
{\em a serious
problem
for the quark model}.

 Negative $\Xi ^0 \to \Lambda \gamma $ asymmetry
  of the (consistent with Hara's theorem) standard
 pole model \cite{Orsay} and ChPT, cannot be obtained in 
 calculations performed with static quarks.
 Such calculations, when applied to NLHDs,
 agree with the expectations based on the
 dominance of CA commutator.
 If experiments find negative $\Xi ^0 \to \Lambda \gamma $ asymmetry,
 the discrepancy with this paper
 cannot be blamed on the
 violation of gauge invariance,
 and would constitute a serious problem {\em for the quark model}.
 A close-to-zero experimental value for
 this asymmetry would constitute
 another problem.

Positive $\Xi ^0 \to \Lambda \gamma $ asymmetry 
does not mean yet that Hara's theorem must be violated, but
that this is likely. If Hara's theorem is satisfied, {\em all} WRHD
amplitudes must vanish in the SU(3) limit. 
Although we cannot vary $m_s-m_d$,
getting experimental information on the $m_s-m_d$
 dependence of theoretical amplitudes
 is to some extent possible.
The experiments in question deal with the sector of weak
p.v.  $\Delta S =0$ nuclear
 transitions and
indicate a lack of the SU(3) suppression for p.v.
 weak couplings of vector mesons to baryons. 
 This hints that there is no such
 suppression for photons either,
 unless some additional
 cancellations occur. 

We have argued that although 
the violation of Hara's theorem is in conflict with
standard theoretical Ans\"atze at hadron level, 
it does not have to be unphysical.
An interpretation of the quark-model result
in terms of an intrinsic baryon nonlocality 
 was presented. 
It was stressed that the issue of Hara's theorem probes the original quark-model
question of apparent quark freedom and additivity at low energy.
It may be that the quark model is an abstraction that went too far
in assigning particle-like properties to quarks, thus leading us
into conflict with the hadron-level description. 
However, it may also be that quark model results should be treated seriously.
Measurement of the $\Xi ^0 \to \Lambda \gamma $
asymmetry will provide crucial information, 
pointing out either to {\em a serious problem for
the quark model} or to {\em intrinsic nonlocality of baryons} when these
are probed by low energy photons.
Whatever the resolution, WRHDs should teach us a lot
about the quark model in the low-energy domain, ie. about the way
quarks combine to form hadrons.

{\bf ACKNOWLEDGEMENTS}\\
I would like to thank A. Horzela and J. Lach for very
useful comments
regarding the presentation of the material of this paper.


\vfill
FIGURE CAPTIONS

Fig.1 Quark diagrams for $W$-exchange contributions to parity-violating
amplitudes of WRHDs.

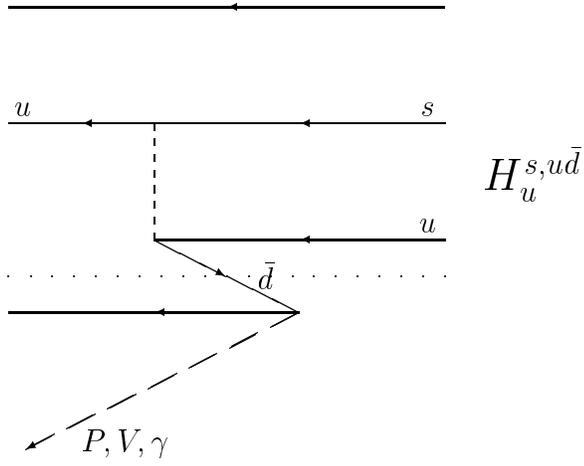
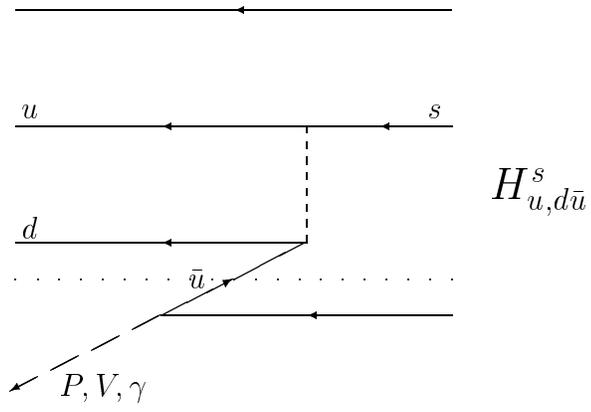
\end{document}